# Gauging Geometry: A didactic lecture

L. Kannenberg[1]
Physics Department
University of Massachusetts Lowell

**ABSTRACT**

Local inertial frame invariance is taken as the fundamental principle of physical geometry, where local inertial frame is represented by a vierbein. Invariance of the vierbein with respect to local Lorentz transformations then expresses local inertial frame invariance. The dynamics of physical geometry develops as a gauge theory of the vierbein that is closely analogous to the Yang-Mills field provided the vierbein connection and curvature correspond to the geometric potential and field respectively. The resulting theory is shown to be equivalent to Einstein's tensor form of relativistic gravitation.

## 1. Introduction

Einstein based "general relativity" at least in part on the requirement that the laws of physics be *generally covariant*, i.e. that they have the same form in all coordinate systems, not just the inertial ones[2]. On this basis he developed his relativistic account of gravitation in 1915. Yet just one year after its final publication[3] Kretschmann[4] showed that *any* theory can be hammered into generally covariant form, and in 1923 Cartan[5] proceeded to do just that for Newtonian gravitation. It took the genius of an Einstein to produce what is arguably the most successful physical theory ever from a physically sterile principle. We mere mortals, however, require a physical foundation for a physical theory. This lecture shows how to implement such a foundation, *local inertial frame invariance*, by applying Noether's theorem[6] on gauge invariance (her "second theorem") while respecting the Master's requirement of general covariance.

---

## 2. The vierbein field.

A frame locally inertial[7] about a typical event $\mathcal{P}$ is characterized by a *vierbein*[8], a quartet of vectors $\boldsymbol{e}_a(\mathcal{P})$, $a$ = 0-3, conventionally orthonormal,

$$\boldsymbol{e}_a(\mathcal{P}) \cdot \boldsymbol{e}_b(\mathcal{P}) = \eta_{ab}, \tag{1}$$

where $\eta_{ab}$ is the Lorentz metric. A *particular* event, $O$, and its vierbein $\boldsymbol{e}_a(O)$ is designated the *lab frame*, with which a coordinate system $\{x^i, i = 0\text{-}3\}$[9] can be constructed that uniquely labels every event accessible from $O$; for example, $x^i(\mathcal{P})$ labels event $\mathcal{P}$[10]. Once coordinate system $\{x^i\}$ is in place countless others, e.g. $\{x'^i\}$, can be generated, where

$$x'^i = x'^i(x) \tag{2}$$

under the usual smoothness conditions; but the lab frame itself remains fixed once and for all. We express general covariance under Eq. (2) by the requirement that the tensor components $v'^i(x'(\mathcal{P}))$ of the vector field $\boldsymbol{v}$ in the $\{x'\}$ coordinates for $\mathcal{P}$ be related to those in the $\{x\}$ coordinates in the usual way,

$$v'^i(x'(\mathcal{P})) = v^j(x(\mathcal{P}))(\partial_j x'^i) \tag{3}$$

and so forth. The *covariant derivative* of $v^i$ is an operation of the form

$$^1D_j v^i = \partial_j v^i + \Gamma^i{}_{jk} v^k \tag{4}$$

that responds to Eq. (2) as a tensor of valence (1,1). Of $\Gamma^i{}_{jk}$ we require only that for any given $\mathcal{P}$ there be a coordinate system $\{x\}$ for which

$$\Gamma^i{}_{jk}(x(\mathcal{P})) = 0. \tag{5}$$

This of course implies that $\Gamma^i{}_{jk}$ must be symmetric under interchange of its covariant indices. Coordinates for which Eq. (5) holds are *geodesic* about $\mathcal{P}$. At this point it is perhaps worth noting that here we take the $\Gamma^i{}_{jk}$, the components $g_{ij}$ of the metric tensor, and the components $R^i{}_{jkl}$ of the curvature tensor to be functions of the components $e_a{}^i$ of the vierbein (and their derivatives); the explicit forms of these functions will be dictated by the formalism.

---

[7] The physical procedures used to *establish* a locally inertial frame, and from it a coordinate system, are specified in an earlier lecture.
[8] "Vierbein", not "tetrad", since the latter excites mathological overtones that I wish to suppress. Vierbein indices are taken from the beginning of the alphabet.
[9] Tensor indices are taken from the middle of the alphabet.
[10] The coordinate system covers a single typical region of the plenum of events. For present purposes it suffices to restrict our attention to just this region.

Since the matrix of vierbein components $e_a{}^i$ is nonsingular it has an inverse with components conveniently written $e^a{}_i$ so that

$$e^a{}_i e_b{}^i = \delta^a{}_b \tag{6}$$

and

$$e_a{}^i e^a{}_j = \delta^i{}_j . \tag{7}$$

Equation (6) validates the use of the Lorentz metric and its inverse to raise and lower vierbein indices, e.g.

$$\eta_{ab} = \eta_{ac}\delta^c{}_b = \eta_{ac} e^c{}_i e_b{}^i = e_{ai} e_b{}^i = \boldsymbol{e}_a \cdot \boldsymbol{e}_b , \tag{8}$$

while Eq. (7) illustrates that the vierbein plays the role of a projection operator between the vierbein and tensor components of a typical vector:

$$v^i = v^a e_a{}^i , \quad v^a = v^i e^a{}_i . \tag{9}$$

The projection of the Lorentz metric,

$$g_{ij} = e^a{}_i \eta_{ab} e^b{}_j , \tag{10}$$

which we recognize as the metric tensor, and its inverse then serve the analogous purpose of raising and lowering tensor indices. Equation (7) also expresses the completeness of the vierbein, i.e. that any vector $\boldsymbol{v}$ on $\mathcal{P}$ can be expanded in terms of its vierbein components,

$$\boldsymbol{v} = v^a \boldsymbol{e}_a . \tag{11}$$

In particular, if $\boldsymbol{'e}_a$ is a quartet of vectors on $\mathcal{P}$ whose expansions in terms of the vierbein $\boldsymbol{e}_a$ on $\mathcal{P}$ are given by

$$\boldsymbol{'e}_a = L^{-1b}{}_a \boldsymbol{e}_b \tag{12}$$

(the notation is conventional), where the components $L^{-1a}{}_b$ satisfy the Lorentz condition

$$L^{-1c}{}_a \eta_{cd} L^{-1d}{}_b = \eta_{ab} , \tag{13}$$

then the '$e_a$ also qualify as a vierbein on $\mathcal{P}$. Since Eq. (13) imposes 10 conditions on the components $L^{-1a}{}_b$, there is a sixfold infinity of vierbeine on each event. For a given vierbein $e_a(\mathcal{P})$ on $\mathcal{P}$ there is consequently on every event $Q$ neighboring $\mathcal{P}$ a vierbein $e_a(Q)$ tht is only infinitesimally different from $e_a(\mathcal{P})$. It is therefore always possible to establish a smooth *field* of vierbeine about $\mathcal{P}$. A local Lorentz transformation $L^a{}_b(\mathcal{P})$ on $e_a(\mathcal{P})$ then necessarily induces a smooth field $L^a{}_b(x)$ of local Lorentz transformations on the field of vierbeine $e_a(x)$ about $\mathcal{P}$. The principle of frame invariance we therefore express as the invariance of the laws of physics to the "geometric gauge group" of local Lorentz transformations on the vierbein field.

### 3. Gauge theory and general covariance.

The archetype of a gauge theory is of course Maxwell's electrodynamics, a massless vector field described by a potential $A_i$ and a field tensor $F_{ij}$. It is invariant under a gauge transformation of the form

$$\Delta A_i = \partial_i \delta\lambda \tag{14}$$

$$\Delta F_{ij} = 0, \tag{15}$$

where $\delta\lambda$ is here taken to be infinitesimal (so that Eq. (14) implies only the algebra of the group; but this suffices for present purposes). For the Lagrangian of the system to be invariant (up to divergences) under this gauge transformation when coupled to a charged source $\psi$ that responds to the gauge transformation according to

$$\Delta\psi = ie\delta\lambda\psi, \tag{16}$$

Noether's second theorem requires replacement of the ordinary derivative of $\psi$ with the (electromagnetic) *gauge derivative*

$$^2D_j\psi = \partial_j\psi - ieA_j\psi \tag{17}$$

whether or not $\psi$ satisfies the field equations. We observe that the electromagnetic field appears in the commutator of two gauge derivatives on $\psi$:

$$[^2D_i, {}^2D_j]\psi = -ieF_{ij}\psi. \tag{18}$$

The field equations themselves,

$$F_{ij} = \partial_i A_j - \partial_j A_i \tag{19}$$

$$\partial_j F^{ij} = j^i(\psi) \,, \tag{20}$$

specify first, the relation between the potential and the field, and second, the "dynamic" field equation.

The Maxwellian gauge group is Abelian. More instructive for our present purposes is Maxwell's descendant the Yang-Mills field, a three-component "isovector" potential $\boldsymbol{b}_i$ and field tensor $\boldsymbol{G}_{ij}$, covariant with respect to the non-Abelian gauge transformation

$$\Delta \boldsymbol{b}_i = g \delta\boldsymbol{\lambda} \times \boldsymbol{b}_i + \partial_i \delta\boldsymbol{\lambda} \,, \tag{21}$$

$$\Delta \boldsymbol{G}_{ij} = g \delta\boldsymbol{\lambda} \times \boldsymbol{G}_{ij} \,. \tag{22}$$

Just as in the Maxwellian case, the potentials "carry" the gauge transformation, while both potentials and fields are subject to a rotation in the internal isotopic space. If the Yang-Mills Lagrangian is to be invariant up to divergences under the gauge transformation of Eqs. (21) and (22), Noether's second theorem demands that it have the form

$$\mathcal{L}_{YM} = -\tfrac{1}{2}\, \boldsymbol{G}^{ij} \cdot (\partial_i \boldsymbol{b}_j - \partial_j \boldsymbol{b}_i + g \boldsymbol{b}_i \times \boldsymbol{b}_j) + \tfrac{1}{4}\, \boldsymbol{G}^{ij} \cdot \boldsymbol{G}_{ij} \,. \tag{23}$$

Further, if the isovector field $\boldsymbol{\psi}$ is coupled gauge invariantly to the Yang-Mills field, ordinary derivatives of $\boldsymbol{\psi}$ must be replaced with the (Yang-Mills) gauge derivative

$$^2D_i \boldsymbol{\psi} = \partial_i \boldsymbol{\psi} - g \boldsymbol{b}_i \times \boldsymbol{\psi} \,, \tag{24}$$

with the field $\boldsymbol{G}_{ij}$ appearing in the commutator of two gauge derivatives,

$$[^2D_i, {}^2D_j]\boldsymbol{\psi} = -g \boldsymbol{G}_{ij} \times \boldsymbol{\psi} \,. \tag{25}$$

The free Yang-Mills field equations obtained by independent variation of Eq. (23) with respect to $\boldsymbol{G}^{ij}$ and $\boldsymbol{b}_i$ respectively

$$\boldsymbol{G}_{ij} = \partial_i \boldsymbol{b}_j - \partial_j \boldsymbol{b}_i + g \boldsymbol{b}_i \times \boldsymbol{b}_j \tag{26}$$

$$^2D_j \boldsymbol{G}^{ij} = 0 \,, \tag{27}$$

the first of which relates the potential to the field, the second being the "dynamic" field equation.

To express these theories in generally covariant form the first step is of course to replace the ordinary derivative of each coordinate tensor with the appropriate covariant derivative. Thus for example $\partial_k \boldsymbol{G}^{ij}$ is replaced with $^1D_k \boldsymbol{G}^{ij}$ and so forth, with one term in $\Gamma^k{}_{ij}$ for each tensor index. There is an additional modification of the Lagrangian that has no impact on the Maxwell and Yang-Mills systems but will be important for the geometric field; for $\mathcal{L}$ must be a coordinate relative scalar of weight + 1 in order that the combination $d^4x\mathcal{L}$ appearing in the action integral responds a scalar to a general coordinate transformation. To achieve this we multiply $\mathcal{L}$ by the factor

$$e^{-1} = \det[e^a{}_i] \,, \tag{28}$$

where from Eq. (10) above one has

$$g = \det[g_{ij}] = -e^{-2} \,. \tag{29}$$

The generally covariant Lagrangian density for the Yang-Mills field therefore reads

$$\mathcal{L}_{YM} = e^{-1} \left\{ -\tfrac{1}{2}\, \boldsymbol{G}^{ij} \cdot ({}^1D_i\boldsymbol{b}_j - {}^1D_j\boldsymbol{b}_i + g\boldsymbol{b}_i \times \boldsymbol{b}_j) + \tfrac{1}{4}\, \boldsymbol{G}^{ij} \cdot \boldsymbol{G}_{ij} \right\} \,, \tag{30}$$

but the $\Gamma^k{}_{ij}$ factors in Eq. (30) are "invisible" since they are symmetric in their lower indices. They are by no means invisible everywhere, however; for example, the generally covariant form of Eq. (27) is

$$\partial_j \boldsymbol{G}^{ij} + \Gamma^i{}_{jk}\boldsymbol{G}^{kj} + \Gamma^j{}_{jk}\boldsymbol{G}^{ki} - g\boldsymbol{b}_j \times \boldsymbol{G}^{ij} = 0 \,. \tag{31}$$

Since the generally covariant form for the gauge derivative of a mixed structure like $\boldsymbol{G}^{ij}$ will be ubiquitous in a generally covariant formalism it is convenient to define the *general derivative* of such a mixed structure, e.g. the gauge and coordinate vector $\boldsymbol{\psi}^i$, as

$$D_j \boldsymbol{\psi}^i = \partial_j \boldsymbol{\psi}^i + \Gamma^i{}_{jk}\boldsymbol{\psi}^k - g\boldsymbol{b}_j \times \boldsymbol{\psi}^i \,. \tag{32}$$

One cautionary note is in order: Since the potentials, $A_i$ for electromagnetism and $\boldsymbol{b}_i$ for Yang-Mills theory, are *not* gauge vectors (although they *are* coordinate vectors) it is inappropriate to apply this notation to them.

### 4. Gauging geometry.

We anticipate that the invariance of the "geometric field" of vierbeine under local Lorentz transformations produces results analogous to those observed in the above examples. An infinitesimal local Lorentz transformation takes the form

$$L^a{}_b = \delta^a{}_b + \varepsilon^a{}_b \,, \tag{33}$$

where to maintain the Lorentz condition to leading order it is necessary that

$$\varepsilon_{ab} = \eta_{ac}\varepsilon^c{}_b = -\varepsilon_{ba} . \tag{34}$$

Under such a transformation the vierbein components $e_a{}^i$ respond according to

$$\Delta e_a{}^i = -\varepsilon^b{}_a e_b{}^i \tag{35}$$

(and similarly for the components of any "local" vector). For the (geometric) gauge derivative of the vierbein, which takes the form

$$^2D_j e_a{}^i = \partial_j e_a{}^i - \omega^b{}_{aj} e^i , \tag{36}$$

to respond to Eq. (33) as a local vector as well, the response of the vierbein connection $\omega^a{}_{bj}$ must be

$$\Delta\omega^a{}_{bj} = \varepsilon^a{}_c \omega^c{}_{bj} - \varepsilon^b{}_c \omega^a{}_{bj} - \partial_j \varepsilon^a{}_b . \tag{37}$$

The vierbein curvature $R^a{}_{bij}$, which appears in the commutator of two gauge derivatives on a local vector,

$$[^2D_i, {}^2D_j] v^a = R^a{}_{bij} v^b , \tag{38}$$

is a local tensor of valence (1,1):

$$\Delta R^a{}_{bij} = \varepsilon^a{}_c R^c{}_{bij} - \varepsilon^b{}_c R^a{}_{cij} . \tag{39}$$

Since the vierbein connection "carries" the geometric gauge transformation in the same way the electromagnetic potential $A_i$ and the Yang-Mills potentials $\boldsymbol{b}_i$ carry their gauge transformations it is natural to interpret $\omega^a{}_{bj}$ as the geometric potential, and similarly to interpret the vierbein curvature as the vierbein field. The vierbein itself we may interpret as a geometric "pre-potential".

The geometric Lagrangian is a function of the pre-potential, potential, and field (and their first derivatives) only. Variation of the Lagrangian is required to produce: [a] the expression for the curvature in terms of the field, the analog of Eq. (19) for the electromagnetic field and Eq. (26) for the Yang-Mills field; [b] the "dynamic" equation for the geometric field, the analog of Eqs. (20) and (25); and [c] the expression for the geometric potential in terms of the pre-potential, which has no analog in the examples given but must be present if the vierbein, as the direct measure of the field of frames, is in fact the fundamental structure of the theory. In addition, the projection property of the vierbein and its inverse must be able to produce the more familiar tensor form of the theory from the gauge theory version.

To satisfy the above requirements the Lagangian must be the sum of two terms,

$$\mathcal{L} = \mathcal{L}_1(R_{abij}, \omega_{abj}) + \mathcal{L}_2(e_a{}^i, \omega_{abj}) , \tag{40}$$

where as the notation suggests $\mathcal{L}_1$ is a function of the potential and field only, the analog of the (generally covariant form of) the electromagnetic and Yang-Mills Lagrangians, and $\mathcal{L}_2$ is a function of the pre-potential and the potential. Noether's second theorem requires that in the Lagrangian the derivative of the potential may only appear with the potential itself in the gauge covariant combination

$$M_{abij}(\omega) = \partial_i \omega_{abj} - \partial_j \omega_{abi} + \omega_a{}^c{}_i \omega_{cbj} - \omega_a{}^c{}_j \omega_{cbi} \tag{41}$$

and therefore Eq. (40) can be written in the more compact form

$$\mathcal{L} = \mathcal{L}_1(R_{abij}, M_{abij}(\omega)) + \mathcal{L}_2(e_a{}^i, M_{abij}(\omega)) . \tag{42}$$

Since $\mathcal{L}_1$ cannot include the vierbein its status as a scalar density must depend on a purely numerical invariant of the proper weight. The only such structure available is the Levi-Civita symbol (not the tensor) $\varepsilon^{ijkl}$; consequentially $\mathcal{L}_1$ must have the form

$$\mathcal{L}_1 = -\tfrac{1}{2} R^{ab}{}_{ij} \varepsilon^{ijkl} M_{abkl}(\omega) + \tfrac{1}{4} R^{ab}{}_{ij} \varepsilon^{ijkl} R_{abkl} . \tag{43}$$

The $\mathcal{L}_2$ term, it will turn out, is the vierbein form of the Einstein Lagrangian:

$$\mathcal{L}_2 = e^{-1}(e^{aj} e^{bi} M_{abij}(\omega) + \lambda) , \tag{44}$$

where $\lambda$ is an undetermined constant. Variation of $\mathcal{L}$ with respect to $R_{abij}$ is straightforward, and produces

$$R_{abij} = M_{abij}(\omega) = \partial_i \omega_{abj} - \partial_j \omega_{abi} + \omega_a{}^c{}_i \omega_{cbj} - \omega_a{}^c{}_j \omega_{cbi} , \tag{45}$$

the geometric analog of Eqs. (19) and (26). Variation of $\mathcal{L}$ with respect to $e_a{}^i$ yields

$$e^{aj} M_{abij}(\omega) - \tfrac{1}{2}(e_{bi} e^{aj} e^{ck} M_{ackj}(\omega) + \lambda) = 0 \tag{46}$$

which is the vierbein version of Einstein's equation, as we shall see. Variation of $\mathcal{L}$ with respect to the potential $\omega_{abi}$ is somewhat more involved. We find

$$\delta_\omega \mathcal{L} = -\tfrac{1}{2}\, {}^2D_k R^{ab}{}_{ij} \varepsilon^{ijkl} \delta\omega_{abl} - {}^2D_i(e^{-1} q^{abij}) \delta\omega_{abj} = 0 \tag{47}$$

up to divergences, where

$$q^{abij} = (e^{aj} e^{bi} - e^{ai} e^{bj}) . \tag{48}$$

If in accordance with Eq. (45) $R_{abij}$ is replaced with $M_{abij}(\omega)$ in Eq. (47), the derivative in the first term vanishes as a consequence of the Bianchi identity

$$^2D_i M_{abjk}(\omega) + {}^2D_j M_{abki}(\omega) + {}^2D_k M_{abij}(\omega) = 0 \tag{49}$$

and therefore this term does not affect the field equations of the theory[11]. The remaining term of $\delta_\omega \mathcal{L}$ then yields

$$^2D_j(e^{-1}q^{abij}) = ({}^2D_j e^{-1})q^{abij} + e^{-1}({}^2D_j q^{abij}) = 0 \ . \tag{50}$$

Upon multiplying Eq. (50) by $q_{abij}$ we obtain

$$^2D_j e^{-1} = e^{-1} e^{dk}\, {}^2D_k e_{dj} \tag{51}$$

which, substituted back into Eq. (50), produces

$$e^{dk} q^{abij}\, {}^2D_k e_{dj} + {}^2D_j q^{abij} = 0 \ . \tag{52}$$

If we lower the indices $a$ and $b$ in Eq. (52) and multiply the result by $e_{ci}$ we get

$$\eta_{bc}(e^{dk}e_a{}^j\, {}^2D_k e_{dj} + {}^2D_j e_a{}^j) - \eta_{ca}(e^{dk}e_b{}^j\, {}^2D_k e_{dj} + {}^2D_j e_b{}^j) +$$

$$+ e_{ci}e_a{}^j\, {}^2D_j e_b{}^i - e_{ci}e_b{}^j\, {}^2D_j e_a{}^i = 0 \ . \tag{53}$$

Equation (53) simplifies considerably, since

$$e^{dk}e_a{}^j\, {}^2D_k e_{dj} = -e^{dk}e_{dj}\, {}^2D_k e_a{}^j = -{}^2D_j e_a{}^j \ , \tag{54}$$

leaving us with

$$e_{ci}e_a{}^j\, {}^2D_j e_b{}^i - e_{ci}e_b{}^j\, {}^2D_j e_a{}^i = 0 \ . \tag{55}$$

It is convenient to rewrite this as

$$e_a{}^i e_b{}^j({}^2D_j e_{ci} - {}^2D_i e_{cj}) = 0 \ , \tag{56}$$

or, upon expanding the gauge derivatives,

$$\Omega_{abc} = \omega_{bca} - \omega_{acb} \ , \tag{57}$$

where

---

[11] See also C. Lanczos, *Ann. Math.* **39**, 842-850 (1938).

$$\Omega_{abc} = e_a{}^i e_b{}^j (\partial_j e_{ci} - \partial_i e_{cj}) \tag{58}$$

is the *object of anholonomy*, and I have written

$$\omega_{abc} = \omega_{abi} e_c{}^i . \tag{59}$$

Taking cyclic permutations of the indices in Eq. (57) produces two further equations like it, the three of which can be combined to yield

$$\omega_{abi} = \tfrac{1}{2} e^c{}_i (\Omega_{cab} + \Omega_{bca} - \Omega_{abc}) , \tag{60}$$

the expression for the geometric potential, that is the vierbein connection, in terms of the "pre-potential", the vierbein, and its derivatives.

The original tensor form of the theory developed by Einstein can be obtained as a kind of projection of the vierbein form. The response of the $\Gamma^k{}_{ij}$ to a coordinate transformation is determined by the requirement that the covariant derivative

$${}^1D_i v^k = \partial_i v^k + \Gamma^k{}_{ij} v^j \tag{61}$$

responds to a general coordinate transformation as a tensor of valence (1,1). In passing it is worth notice that since this requirement is completely different from that of local Lorentz covariance, $\Gamma^k{}_{ij}$ cannot be the projection of the vierbein connection $\omega_{abi}$. Noether's second theorem requires that $\Gamma^k{}_{ij}$ and its derivatives appear in the Lagrangian only in the combination

$$R^k{}_{lij}(\Gamma) = \partial_j \Gamma^k{}_{il} - \partial_i \Gamma^k{}_{jl} + \Gamma^k{}_{jm} \Gamma^m{}_{il} - \Gamma^k{}_{im} \Gamma^m{}_{jl} \tag{62}$$

which of course we recognize as the tensor form of the curvature tensor. Einstein's Lagrangian for this system reads

$$\mathcal{L} = (-g)^{1/2} (g^{ij} R^k{}_{ijk}(\Gamma) + \lambda) , \tag{63}$$

that is, it is already reduced from the two-term tensor analog of Eq. (42). Variation of Eq. (63) with respect to $g^{ij}$ results in Einstein's famous equation,

$$R^k{}_{ijk}(\Gamma) - \tfrac{1}{2} g_{ij} (g^{kl} R^m{}_{klm}(\Gamma) + \lambda) = 0 \tag{64}$$

the tensor analog of Eq. (46). Further, the variation of Eq. (63) with respect to $\Gamma^k{}_{ij}$ yields the equation

$$0 = {}^1D_k\big((-g)^{1/2} g^{ij}\big) = g^{ij} \, {}^1D_k (-g)^{1/2} + (-g)^{1/2} \, {}^1D_k g^{ij} \tag{65}$$

the tensor analog of Eq. (50). With a little algebra this can be reduced to

$$^1D_k g^{ij} = 0 \,. \tag{66}$$

This is easily rearranged to produce the familiar form

$$\Gamma^k{}_{ij} = \tfrac{1}{2} g^{kl}(\partial_i g_{jl} + \partial_j g_{li} - \partial_l g_{ij}) \,, \tag{67}$$

the tensor analog of Eq. (60), and validates the identification of $\Gamma^k{}_{ij}$ as the Christoffel symbol $\{{}_i{}^k{}_j\}$.

Equation (66) ensures that raising and lowering tensor indices with $g_{ij}$ and its inverse commute with the covariant derivative $^1D_j$, just as raising and lowering vierbein indices with $\eta_{ab}$ and its inverse commutes with the (geometric) gauge derivative $^2D_j$. The projection of tensor indices onto vierbein indices and *vice versa* does not however commute with either type of derivative as a consequence of Eq. (56) and the vierbein form of Eq. (66), i.e.

$$e^a{}_i \,{}^1D_k e_{aj} + e_{aj} \,{}^1D_k e^a{}_i = 0. \tag{68}$$

On the other hand, the tensor/vierbein projection operator $e_a{}^i$ *does* commute with the (geometric) *general derivative* defined in analogy with the (Yang-Mills) general derivative of Eq. (32), i.e.

$$D_j \psi_a{}^i = \partial_j \psi_a{}^i - \omega^b{}_{aj} \psi_b{}^i + \Gamma^i{}_{kj} \psi_a{}^k \,, \tag{69}$$

since it is easy to show that

$$D_j e_a{}^i = 0 \tag{70}$$

holds if and only if Eqs. (50) and (67) are satisfied.

Coupling to other fields is, as with other gauge theories, effected through the general derivative, and therefore the interaction Lagrangian will be a function of $e_a{}^i$, $\omega_{abj}$, and $\Gamma^k{}_{ij}$ (and never their derivatives). It is interesting that the putative geometric *field*, that is $R^a{}_{bij}$, would only be present if a contribution including it or its relatives such as the Weyl tensor were inserted *ad hoc*, and therefore can be excluded on "minimal coupling" grounds.

## 5. Concluding remarks.

This is hardly the first formulation of "general relativity" in terms of the vierbein, which was introduced by Weyl[12] to accommodate electrons and other half-integer spin particles in a generally covariant field theory. Einstein himself took advantage of the vierbein in his search for a unified field theory[13]. It is also not the first to treat relativistic gravitation as a gauge theory analogous in some sense to that of the Yang-Mills field[14]. Attempted here is simply a straightforward development of the theory taking the vierbein as fundamental from the outset, and not just as an adjunct to the tensor form of the theory.

---

[12] H. Weyl, *Zeitschr. d. Phys.* **56**, 330-352 (1929); see also F. A. Kaempffer, *Phys. Rev.* **165**, 1420-1423 (1968).

Since however the present formalism can be reduced to Einstein's more familiar tensor version (in the absence of spinor fields), is this all merely an academic exercise, "a tale told by an idiot, full of sound and fury, signifying nothing"? I think not. It has at least the minimum virtue of providing a sounder physical underpinning for Einstein's theory than the shaky requirement of general covariance. In addition, re-expressing "physical geometry" in terms of local inertial frame invariance brings it within the ambit of gauge theory rather than being a formalism *sui generis*. Moreover, many mathematical techniques have been developed for dealing with Yang-Mills theory and its descendants that are likely to be applicable to this version of Einsteinian gravitation as well. Finally, although the $\mathcal{L}_1$ component of the Lagrangian plays no dynamic role in the classical field theory its presence may nevertheless have consequences for the canonical analysis of the system and its quantization.

---

[13] A. Einstein, *Revs. Mod. Phys.* **20**, 35-39 (1948) and J. Yepez, arXiv 1106.2037.

[14] A fairly comprehensive review is provided by M. Blagojevic and F. Hehl, *Gauge Theories of Gravitation: a reader with commentaries*, London 2013.7ytg